\documentclass[aps,twocolumn, superscriptaddress]{revtex4-1} 

\usepackage{amsmath}  
\usepackage{amsfonts} 
\usepackage{graphicx} 
\usepackage[english]{babel}
\usepackage[dvipsnames]{xcolor}
\usepackage{xfrac} 
\usepackage[normalem]{ulem}
\usepackage{xcolor}
\usepackage{siunitx}
\usepackage{verbatim}

\usepackage{soul}

\providecommand{\keywords}[1]
{
  \small	
  \textbf{\textit{Keywords---}} #1
}

\begin{document}
\title{Clones of the Unclonable: Nanoduplicating Optical PUFs and    Applications} 


\author{E. Marakis$^*$}
\affiliation{MESA+ Institute for Nanotechnology,
University of Twente, PO Box 217, 7500AE Enschede, The Netherlands}
\thanks{These two authors contributed equally, sharing first authorship.}

\author{U.\ R\"{u}hrmair$^*$}
\affiliation{LMU, München, Germany, and University of Connecticut, Storrs, USA}

\author{M.\ Lachner}
\affiliation{LMU, München, Germany}

\author{R.\ Uppu}
\affiliation{MESA+ Institute for Nanotechnology,
University of Twente, PO Box 217, 7500AE Enschede, The Netherlands}
\affiliation{Department of Physics \& Astronomy, University of Iowa, 205 N Madison St, Iowa City, IA 52242, USA}

\author{B.\ \v{S}kori\'{c}}
\affiliation{Eindhoven University of Technology, PO Box 513, 5600 MB Eindhoven, The Netherlands}

\author{P.\ W.\ H.\ Pinkse}
\affiliation{MESA+ Institute for Nanotechnology,
University of Twente, PO Box 217, 7500AE Enschede, The Netherlands}

\date{\today}
\begin{abstract}
Physical unclonable functions (PUFs), physical objects that are practically unclonable because of their andom and uncontrollable manufacturing variations, are becoming increasingly popular as security primitives and unique identifiers in a  fully digitized world. 
One of the central PUF premises states that both friends and foes, both legitimate manufacturers and external attackers alike, cannot clone a PUF, producing two instances that are the same.  
Using the latest nanofabrication techniques, we show that this premise is not always met:  We demonstrate the possibility of effective PUF duplication through sophisticated manufacturers by producing 63 copies of a non-trivial optical scattering structure which exhibit 
essentially the same scattering behavior.  
The remaining minuscule differences are close to or below 
noise levels, whence the duplicates have to be considered fully equivalent from a PUF perspective.  The possibility for manufacturer-based optical PUF duplication has positive and negative consequences at the same time:  While fully breaking the security of certain schemes, it enables new applications, too.  For example, it facilitates unforgeable labels for valuable items; the first key-free group identification schemes over digital networks; or new types of encryption/decryption devices that do not contain secret keys.

\end{abstract}


\maketitle 

\section{Introduction}

Physical unclonable functions (PUFs) have exerted a major influence on cryptography and hardware security over the last two decades \cite{Pappu2002,Gassend2002,Her2014,Ruh2014Glance,Gao2020,Lugli,Csaba}. Among others, PUFs can remove the need to store secret keys
in vulnerable hardware \cite{Skoric2012, Goorden2014, Ruh2022}; form tamper-detecting capsules around electronic circuits to  protect these against adversarial physical access \cite{Gassend2003, GaborTransf, Imm2018}; or serve as universal cryptographic primitives, enabling fundamental protocols such as remote identification \cite{Pappu2002}, 
key exchange \cite{Ruh2013,Ruh2010}, or oblivious transfer \cite{Ruh2013,Ruh2010}.   

Since the beginnings, one central PUF promise had been that their uncontrollable manufacturing variations make them practically unclonable for {\it both} external adversaries {\it and} for their original manufacturer {alike}.    
With respect to manufacturers, this decisive property has often been termed {\it ``manufacturer resistance''} in the literature \cite{Ruh2009, Mae2010}.  This feature is obviously essential for the security of most PUF applications, such as the abovementioned  PUF-based identification protocols \cite{Pappu2002}.  
It also strictly demarcates PUF hardware from traditional systems containing digital secret keys:  Once such keys are known to manufacturers, they can easily be duplicated in other hardware, while PUFs could not --- at least so the narrative went \cite{Pappu2002,Gassend2002,Her2014,Ruh2014Glance,Gao2020}.   

We here show for the first time that this central PUF premise of manufacturer resistance can be broken by the latest  fabrication methods.  To this end, we demonstrate the effective duplication of  complex nontrivial optical scattering structures or PUFs through application of high-resolution nanoscale 3D printing by the original PUF manufacturer.  The multiple structures fabricated by us are, and remain, unclonable for external adversaries who do not know their internal features and inner construction plan.  But these structures all possess essentially the same optical scattering behavior (or challenge-response pairs in PUF speak), and thus constitute practically identical, manufacturer-generated PUF duplicates.  
We stress in this context that the word {\it ``cloning''} will be reserved in this manuscript for external adversaries, while the term {\it  ``duplication''} refers to a malicious or honest original manufacturer of a PUF who tries to produce more than one practically identical copy. 

The new capabilities of PUF duplication have various security consequences, some advantageous, some detrimental. These will be detailed in Sec. \ref{sec:consequences}. In the  upcoming Sec. \ref{sec:Creation} we will discuss our nanofabrication method before discusssing the optical quality of the produced samples in Sec. \ref{sec:Quality}.

\section{Creating PUF Duplicates by Direct Laser Writing}
 \label{sec:Creation}

\begin{figure}[t!]
    \centering
    \includegraphics[width=0.5\textwidth]{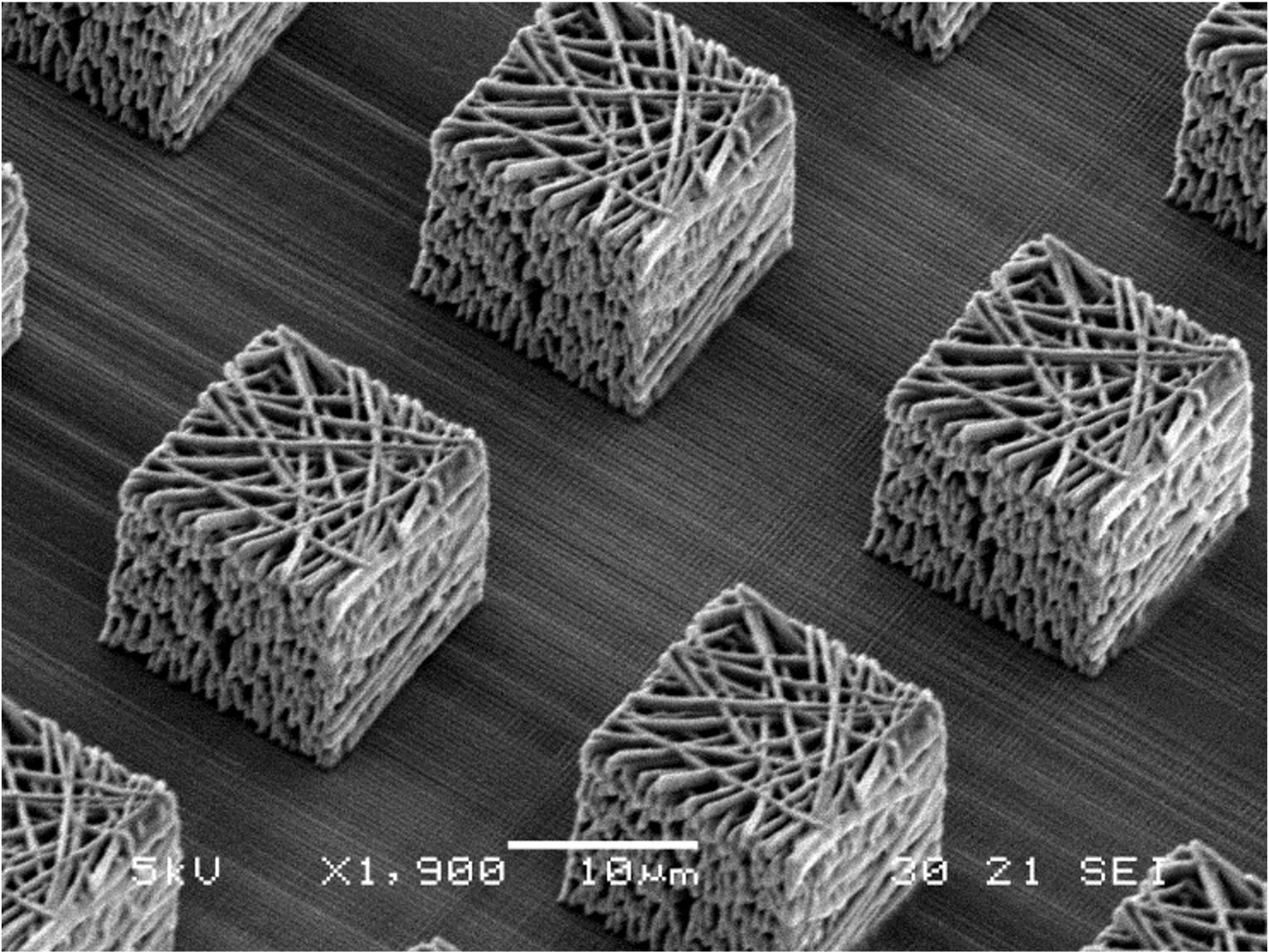}
    \caption{Electron microscope image of a set of  multiple-scattering samples made by high-resolution direct laser writing. The $(15\,{\rm \mu m})^3$ polymer structures are effectively identical in their scattering behavior or, in PUF parlance, in their challenge-response pairs (CRPs).}
    \label{fig:SEMimage}
\end{figure}

Our nanoduplication approach is based on the high-resolution direct laser writing technique described by Marakis {\it et al.} \cite{Marakis2020}. We start with choosing a design (i.e., a digital construction plan) for an optical PUF instance and for all its future duplicates.  In  our case, the design is based on generating a random set of line pieces (``chords'') with an on-average homogeneous density, filling a cubic volume \cite{Marakis2019}. This cubic volume of chords is nanofabricated in a polymer employing a high-resolution 3D printing technique, namely two-photon direct laser writing \cite{Sun2004,Deubel2004,Haberko2013opex,Fischer2013,Farsari2010,Seniutinas2018} (see also Methods). For ease of manufacturing, our design for now features straight lines \cite{Utel2019,Wiersma2013} instead of other, more elaborate possible structures, such as a network of rings or polygons \cite{Haberko2013pra,Haberko2013opex,Muller2014,Haberko2020}. The size of the cubic volumes of $(15\,{\rm \mu m})^3$ was chosen large enough to be in the multiple-scattering regime, and at the same time small enough to prevent systemic imperfections such as cracks and pyramidal distortion \cite{Zhou2015,Bauhofer2017}. In order to construct duplicates that not only possess identical design geometry, but also exhibit near-identical optical behavior in measurements, it is also imperative to maintain experimental parameters such as laser power, scan speed, and age of the polymer identical during the manufacturing process. In practice, we could achieve this only by manufacturing multiple duplicates at the same time in the same production run. Some examples of the resulting, nominally identical nanostructures (or PUF duplicates) are shown in Fig.~\ref{fig:SEMimage}. Overall, we fabricated 99 duplicates, of which 36 were visually damaged. Detailed analysis was carried out on the remaining 63 duplicates for quantifying the similarity using optical readout methods discussed in Sec. \ref{sec:Quality}.

\begin{figure}[t!]
\vspace{-1.5ex}
    \centering
    \includegraphics[width=0.5\textwidth]{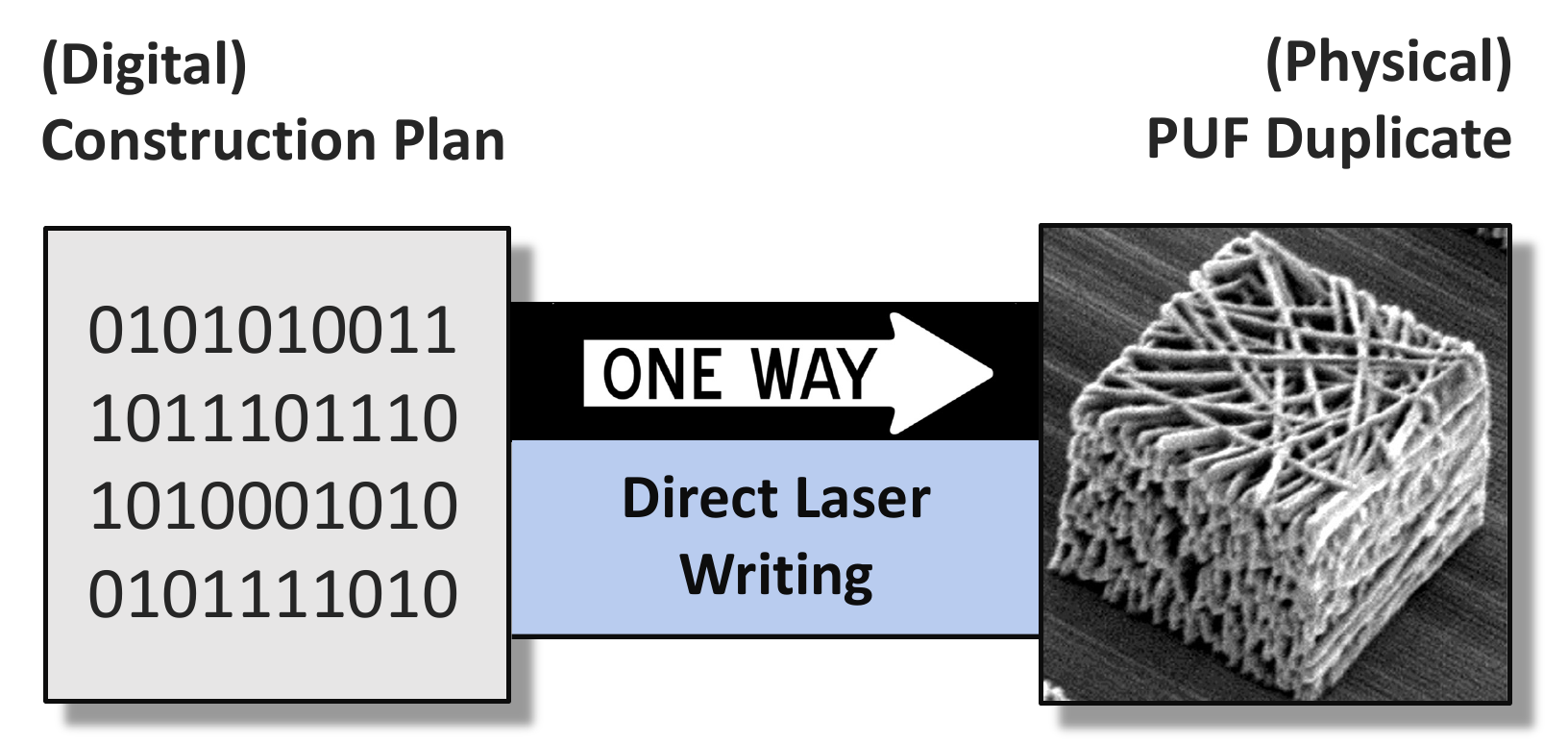}
    \caption{Fabricating a PUF duplicate from an inner construction plan could be seen as a {\it ``one-way process''}:  External adversaries practically cannot obtain or derive  the construction plan from physically inspecting the 3D duplicate. Still, the fabrication process is repeatable for the manufacturer who knows the construction plan, leading to near-identical duplicates, as demonstrated in this paper.}
    \label{fig:One-Way}
\end{figure}

From a cryptographic perspective, the fabrication of one (or more) PUF duplicates from the same construction plan could be seen as a {\it ``physical one-way process''}, as illustrated in Fig. \ref{fig:One-Way}:  Using the same construction plan, many essentially identical PUF duplicates can be fabricated efficiently.
But given the duplicates, external adversaries will find it practically impossible to deduce the exact inner construction plan using either invasive or non-invasive methods.
Invasive methods (dicing, etc.) for reconstructing PUFs are incompatible, since laser-written polymer PUFs are readily damaged by the dicing tool’s mechanical and thermal stress.
This can be alleviated using non-invasive reconstruction methods such as ptychography or diffraction tomography techniques. 
While X-ray light probes, available at synchrotron and free-electron laser facilities, provide high spatial resolution (10 nm), the estimation of refractive index has significant errors $\sim$10\% \cite{Gureyev2013jap}. 
This large estimation error in combination with the sensitivity of speckle correlations (multiple-scattering media have demonstrated sensitivity to $<$1\% variation in the refractive index \cite{Tran2020sens}) disallows the exact reconstruction of the PUF.

An alternative attack strategy is to replicate the optical response (scattering matrix) of the PUF.
Replicating the optical PUF response requires two steps: 1) construction of the optical PUF response and 2) replicating the scattering matrix.
The former can readily be achieved using non-invasive techniques such as optical phase conjugation. In a protocol like QSA \cite{Skoric2012, Goorden2014} this information is even provided as public data. 
However, replicating the large scattering matrices of 3D multiple-scattering PUFs faces challenges.
Firstly, one could use the same method we employed (direct laser writing).
This technique has been successfully demonstrated for fabricating 2.5D structures, such as metasurfaces \cite{Park2020advmat}.
Extending this technique to mimic 3D structures poses challenges due to the non-unique solutions in the inversion problem, which remains an active area of research.
To date, only periodic media that offer additional constraints on the solution search due to the symmetries in the structure have been successfully reconstructed \cite{Freyman2010advfuncmat}.
Instead of attempting to recreate a 3D scattering structure, the rapid advances in reconfigurable photonic integrated circuits offer an alternate pathway for realizing arbitrary scattering matrices.
While exciting, this approach would hit a fundamental limit on the size of the scattering matrix ($\approx 300\times300$ modes) imposed by the scaling in material absorption \cite{Taballione2019opex}.
In comparison, even the tiny $(15 {\rm \mu m})^3$ scattering medium that we employ in this work already has $15/0.63=566$ optical modes and hence a scattering matrix of $566^2$ elements. Further scaling would put the resulting matrix yet further out of reach for existing methods.

\section{Quality and Exactness of Our Duplicates}  \label{sec:Quality}

Although our fabrication technique is 
deterministic, minuscule fluctuations will still cause nominally identically PUF duplicates to differ slightly.  The quality of duplicates is determined by how strongly their responses (i.e., optical scattering images) differ.  We will therefore quantify the likeliness in response images by two classical metrics from the PUF literature in the following, namely the Pearson correlation coefficient and the Hamming distance after application of the Gabor transformation \cite{GaborTransf}.

In order to quantify the similarity we first use the Pearson correlation coefficient
\begin{equation}
    r(I_{\rm A},I_{\rm B})=
    \frac{\langle I_{\rm A} I_{\rm B}\rangle-\langle I_{\rm A}\rangle \langle I_{\rm B}\rangle}
    {\sqrt{\left( \langle I_{\rm A}^2\rangle-\langle I_{\rm A}\rangle^2\right)\left( \langle I_{\rm B}^2\rangle-\langle I_{\rm B}\rangle^2\right)}},
\end{equation}
where $I_{\rm A}$ is the intensity pattern of speckle pattern A and $I_{\rm B}$ is the intensity pattern of speckle pattern B. 
The notation $\langle\cdots\rangle$ stands for spatial
averaging over the entire image. For precise computation of the cross correlation term $\langle I_{\rm A} I_{\rm B}\rangle$ the two patterns are shifted with respect to each other and the maximum value of the image-averaged cross correlation is taken.
The assumption is that the value found here for a particular illumination is representative for an average correlation from a set of randomly chosen illumination patterns as done in the protocol by Goorden {\it et al.} \cite{Goorden2014} and of which the security was analysed by {\v{S}}kori{\'c} {\it et al.} \cite{Skoric2013}.

\begin{figure*}[hbt!]
    \centering
    \includegraphics[width=\textwidth]{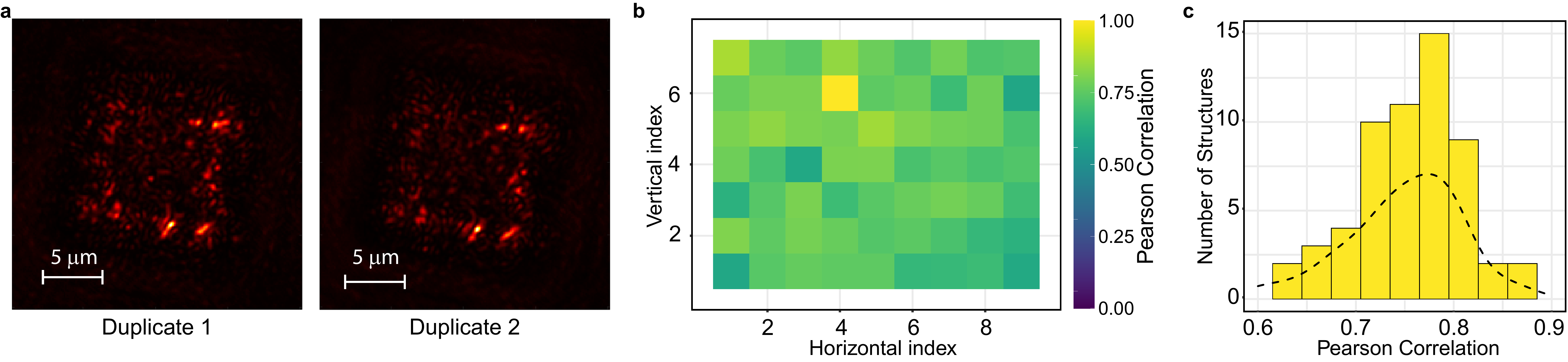}
    \caption{
    (a) Optical microscope images of laser (wavelength = 633 nm) transmission through two duplicates of the nanofabricated scattering structure exhibit very similar speckle patterns. 
    (b) The Pearson correlation of speckle patterns between 63 nominally identical copies of a nanofabricated multiple-scattering sample. 
    The horizontal and vertical axes are horizontal and vertical indices of the structures as they are arranged on one and the same substrate. 
    The yellow square shows the correlation of the one sample that was chosen as reference for this graph, and hence shows its autocorrelation normalized to unity. 
    (c) The histogram of the measured pairwise intensity speckle Pearson correlation values is shown here. 
     The dotted curve represents the estimated probability density function of the Pearson Correlation distribution. It was computed via the kernel density estimation (KDE) using the values retrieved from the samples of our experiments.
    }
    \label{fig:bargraph}
\end{figure*}

We illuminate the substrates with a plane wave and measure the intensity output patterns. Of the employed 63 copies, all mutual cross-correlation coefficients were determined. Two copies which showed the highest cross-correlation with another were chosen as the two PUF duplicates under consideration. The cross correlation of one of this pair with all the 62 other copies is shown in Fig.\,\ref{fig:bargraph}. The histogram peaks around a cross correlation of about 77\,\% and the highest value we found is 87\,\%. 

The probability for an impostor PUF to pass QSA verification
essentially equals the average fidelity between a (complex-valued) response wavefront produced by the impostor PUF and the correct wavefront.
Instead of reporting mutual fidelity values for our scattering structures we present intensity correlations;
in practice this is equally useful, since high intensity correlation implies high fidelity.
This is seen as follows. 
Consider two structures manufactured using the nanofabrication technique from Ref.\cite{Marakis2020}.
Let HI be the event that their intensity correlation exceeds some threshold $T_{\rm I}$.
Let LF be the event that their mutual fidelity is lower than some threshold $T_{\rm F}$.
It holds that
${\rm Prob}[{\rm LF}|{\rm HC}] $
$= \frac{{\rm Prob}[{\rm LF}]\cdot{\rm Prob}[{\rm HC}|{\rm LF}]}{{\rm Prob}[{\rm HC}]}$
$\leq \frac{{\rm Prob}[{\rm HC}|{\rm LF}]}{{\rm Prob}[{\rm HC}]}$.
The ${\rm Prob}[{\rm HC}]$ is significantly larger than zero, since we 
observe some large correlations.
The ${\rm Prob}[{\rm HC}|{\rm LF}]$, however, is exponentially small, since
in a complicated multiple-scattering system
a low fidelity is typically caused not only by differing phases but also by differing
amplitudes.

The speckle patterns under plane-wave illumination of our two record-similar duplicates are depicted in Fig.\ref{fig:bargraph} a).  
They show a similarity of $87\,\%$. This means that a quantum secure authentication experiment like that by Goorden {\it et al.} will accept this duplicate as genuine if the threshold for acceptance of a original PUF is set below that value, even in the case of a large number of CRPs being tested to reduce statistical uncertainty. Given the required robustness against measurement noise, deviations caused by PUF alignment and environmental drifts, we estimate such a high threshold is unlikely to be practical. 
Hence we have, for the first time, demonstrated a non-trivial realization of duplicate PUFs.

To further consolidate above results, we include a similar analysis using the \textit{Fractional Hamming Distance} (FHD), which is defined as the  number of differing bits between two bit strings, divided by their overall length. To retrieve such a string from a speckle pattern, we follow the steps first formulated by Pappu {\it et al.} \cite{Pappu2002}: We apply a Gabor transformation on the speckle pattern, digitize the real part of the complex-transformed image using a threshold of 0, and flatten the resulting binary image into a 1D array of bits. While this approach poses only one of many possible digitization processes, the Gabor transformation is the most prevalent technique in the context of optical PUFs.

Since the Gabor transformation is deterministic, the bit strings retrieved from two identical speckle patterns will be identical as well, so the resulting FHD will be exactly $0$. On the other hand, the expected value of the FHD of two random patterns will be $0.5$, which corresponds to the expected ratio of different bits for two randomly generated binary sequences. All other values indicate a correlation between the speckle patterns. 

In the context of optical PUFs, the FHD is usually established for two different sets. The first one is the \textit{like distribution}, which shows the FHDs between speckle patterns that are supposed to be similar or identical. This usually corresponds to multiple responses retrieved from the same challenges at different points in time. In our case, we use this distribution to evaluate the quality of our duplicate PUFs by computing the FHD between the responses of each copy. The other set is the \textit{unlike distribution}, which corresponds to speckle patterns of scattering objects that are supposed to be very different. This matches the responses of different challenges, which should ideally show no correlations. In our context, this distribution is established by comparing the speckle patterns across different sets of PUF duplicates.

Ideally, the like distribution is centered at $0$ and the unlike distribution at $0.5$, without any variance on either one. However, due to noise in the system and physical correlations in the scattering process, such values are almost never found in practice. In fact, it is not even necessary - it is already sufficient to have two non-overlapping distributions (or at least a sufficiently small overlap), such that a threshold can be defined on which the distributions are separated. By doing so, for any arbitrary pair of speckle patterns, the FHD can be used to determine from which distribution they originated and, therefore, if they are regarded as the same or different responses.

\begin{figure}
    \centering
    \includegraphics[width=0.5\textwidth]{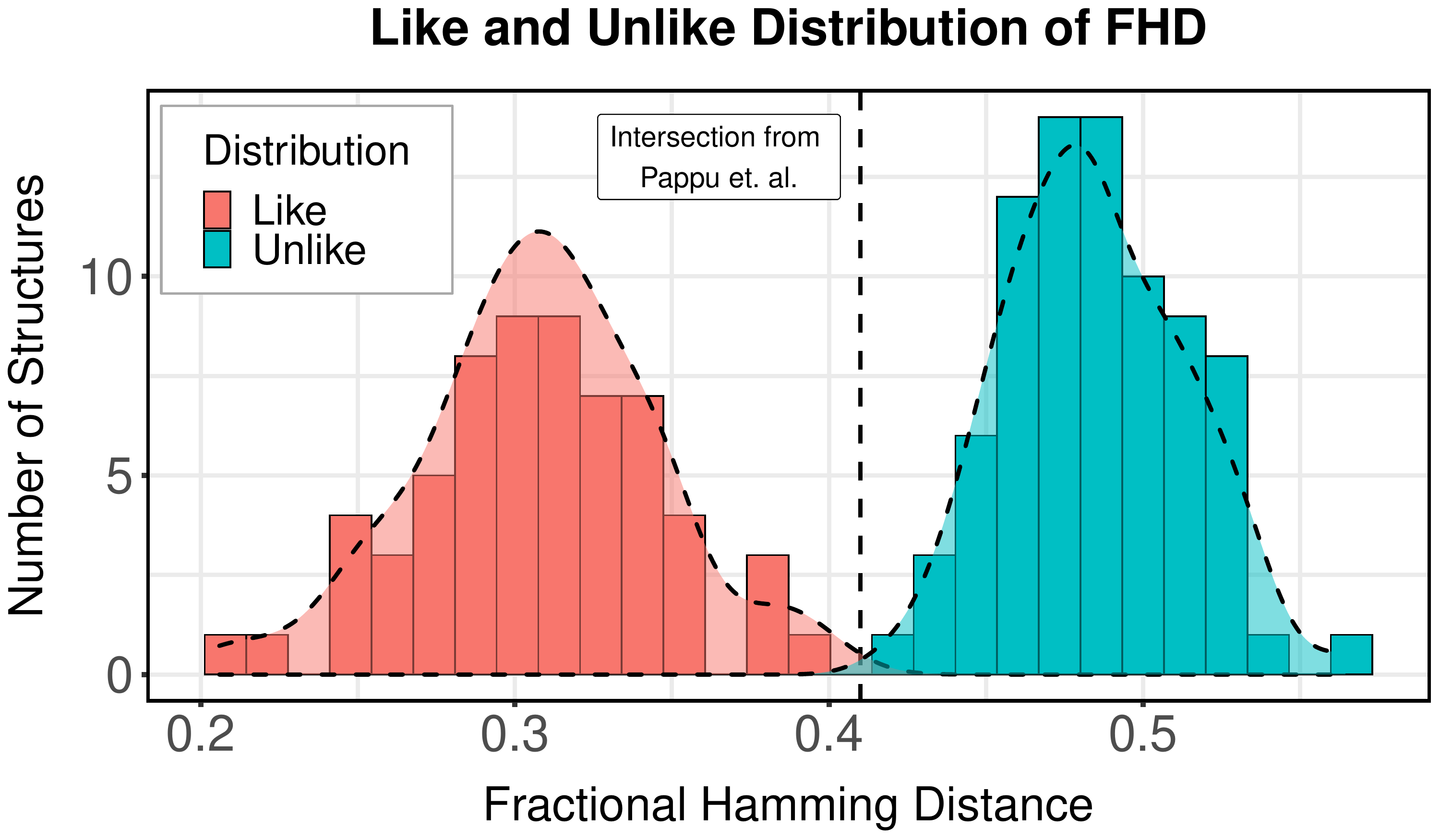}
    \caption{The like and unlike distributions of 62 copies relative to one that was chosen as the ``original'' PUF. There is a clear separation between both distributions at a FHD of $0.4$, which can be used to differentiate whether or not two samples are from the same set of PUF duplicates. The dotted curve represents the estimated probability density function of the like and unlike distributions, computed via the KDE.}
    \label{fig:fhd_dists}
\end{figure}

To estimate the distributions, we follow the same approach from the Pearson correlation analysis before. Additionally, to remove inexpressive correlating bits originating from the shared dark area around each speckle pattern, we first rotate all images counter-clockwise by $\ang{10}$ to align the pattern with the axes and then crop the center part of $125 \times 125$ pixels, which we found to be a fitting value to remove the dark area without losing information from the pattern itself. The resulting distributions can be found in Fig.~\ref{fig:fhd_dists}. We find that the mean FHD of the unlike distribution lies at $0.48$, the mean FHD of the like distribution at $0.3$ and the intersection of both distributions at $0.4$. There is no overlap between both distributions for our samples, so we can clearly distinguish if two speckle patterns are from the set of PUF duplicates. In their original publication, Pappu {\it et al.} \cite{Pappu2002} report approximately the same characteristic values of the distributions, except for a slightly lower like distribution mean of $0.25$. Note, however, that in their experiments, they compare speckle patterns retrieved from the exact same physical system at different moments in time, while we compare duplicated, but physically different systems.

\section{Employing PUF Duplicates for Codemaking and Codebreaking}  \label{sec:consequences}  \label{sec:applications}

\begin{figure}[t!]
    \centering
    \includegraphics[width=0.47\textwidth]{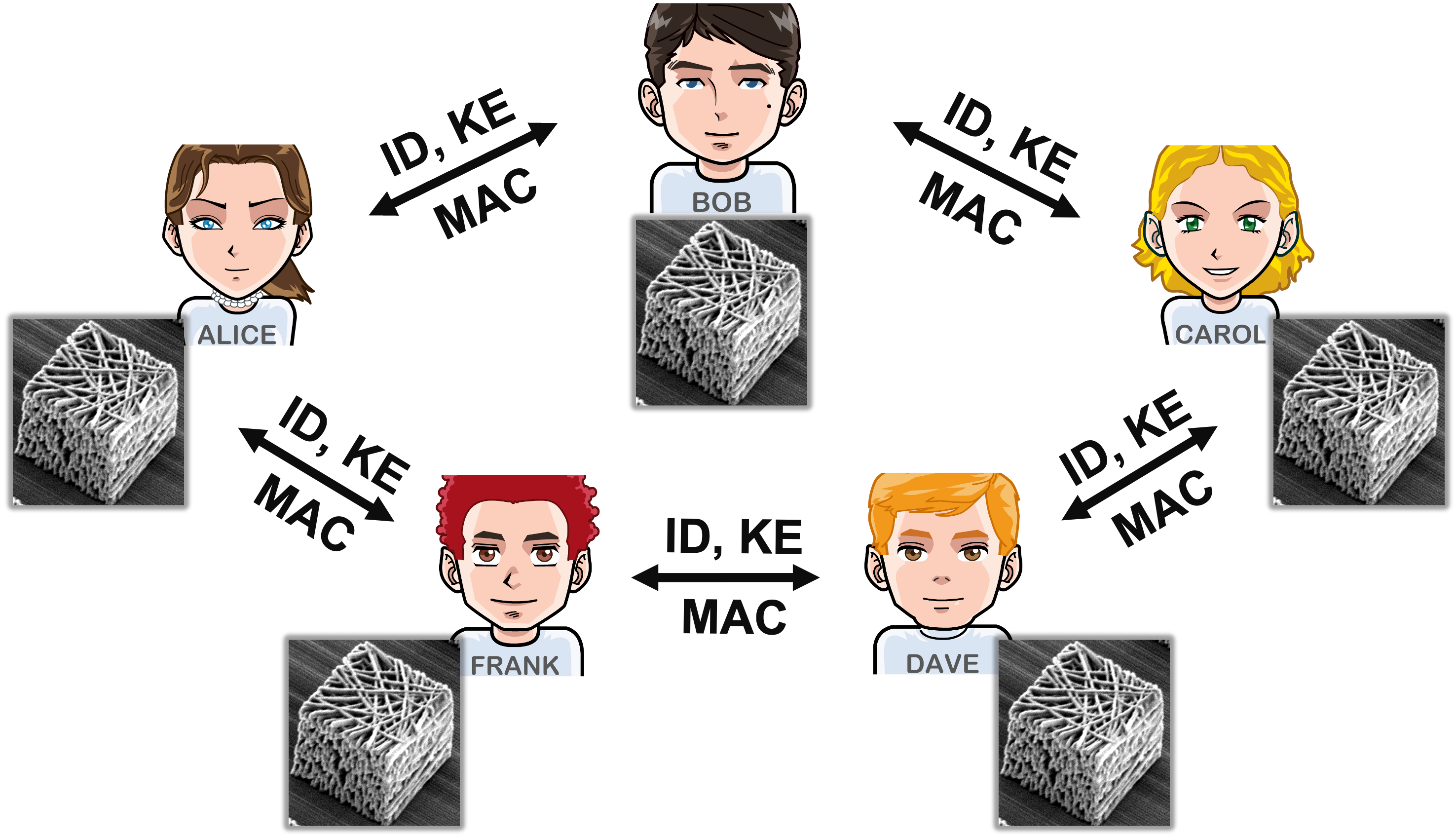}
    \caption{Schematic illustration of mutual identification (ID), key exchange (KE) and message authentication (MAC) among a group of participants in a digital communication network, where each legitimate group member holds an identical PUF duplicate.  No digital cryptographic keys or secret CRP lists need to be stored permanently at the parties, boosting security against hardware and malware attacks.}
    \label{fig:Network}
\end{figure}

The possibility of PUF duplication opens up new avenues for both honest users and attackers.  Starting with the latter, PUF duplicates allow malicious manufacturers to act as hidden adversaries, breaking various PUF schemes without being noticed.  This obviously includes remote identification protocols \cite{Pappu2002}, where manufacturers can cheat on users by keeping a PUF duplicate at production time,  
or by selling such a duplicate to the highest bidder.  Considering typical  applications, such as PUF-based car keys, house keys, passports, or bank cards, can illustrate the massive consequences of this type of attack.  

Another potential fraud scenario concerns advanced cryptographic protocols such as PUF-based key exchange \cite{TS,PhysicalTM,Brzuska} or oblivious transfer \cite{Ruh2010,Brzuska}.  In these schemes, malicious manufacturers can almost trivially cheat on their communication partners by unnoticedly using PUF duplicates instead of ``real'' PUFs.  
Let us first illustrate this type of attack by the example of PUF-based oblivious transfer (OT).  OT is the probably most fundamental single cryptographic protocol, upon which essentially any other cryptographic task can be built \cite{Kilian}. 
Please recall that all existing PUF-based OT protocols (such as \cite{Ruh2010,Ruh2013,vanDijk2012,Ostrovsky2013}) assume that the so-called {\it ``OT sender''} holds a PUF at the protocol start, subsequently measures a subset of all CRPs, and then physically passes the PUF to the second protocol participant, the {\it ``OT receiver''}.  The purpose of this physical transfer is that the sender subsequently can no longer measure or deduce arbitrary responses of his choice from the PUF, as it is now in someone else's physical possession. 
This restriction can easily be violated and overcome via PUF duplication, however.  The sender simply uses several duplicates instead of a ``normal'' PUF, and keeps one PUF duplicate while sending away the other.  
This subsequently allows him to cheat in the OT-protocol, concretely to deduce which of the two OT-bits was learned by the receiver and which was not.  
From a technical perspective, the resulting attacks are similar to the ones presented in \cite{Ruh2013} for the case of clonable bad PUFs.  In passing, this renders our optical PUF duplicates the first practical demonstration of a non-trivial clonable bad PUF, affirmatively answering a decade-old question from \cite{Ruh2013}. 

Similar observations hold for PUF-based cryptographic key exchange (KE) \cite{TS,PhysicalTM,Brzuska}:  Assuming that the same Strong PUF is employed in several consecutive KE protocols between different users, a malicious manufacturer again can cheat easily.  To this end, he produces several PUF duplicates (instead of a single, ``real'' PUF) and keeps at least one of them in his hands.   
He then eavesdrops all challenges $C_1, \ldots, C_k$ that are employed for key derivation in the abovementioned consecutive KE protocols.  Recall that these are sent between users in the clear \cite{TS,PhysicalTM,Brzuska}.  He subsequently applies these challenges to his local PUF duplicate (instead of the original PUF), obtains the same responses as users, and successively derives the same secret keys, breaking the protocol.   
The full attack and further comments will be detailed later.

Fortunately, PUF duplication also has various positive and constructive applications:  
Consider, as a first example, a group of users, all equipped with their local, identical PUF duplicate.  This setup allows any user to identify as group member to all other group members, and to do so remotely via digital communication in a standard network.  
To this end, one user (acting as {\it ``verifier''} for the group membership) chooses several PUF challenges at random, and sends them to the second user (the {\it ``prover''}, who wants to demonstrate his group membership).  The prover then applies the challenges to his PUF duplicate and returns the obtained responses to the verifier via the digital network.  The verifier subsequently applies the same challenges to his local PUF duplicate, and compares his responses to those received from the prover.  If they match within a certain error threshold, the prover is considered successfully identified as group member.  
Similar schemes can be designed for the authentication of arbitrary messages between all group members (compare Protocol 3 of \cite{SIMPL}), establishing a group-internal authenticated channel. 

This authenticated group channel and the possession of the duplicates can also readily be employed for key exchange protocols, in which all group members learn the key, while external adversaries will not.  To this end, randomly chosen challenges are sent on the authenticated group channel, and users derive keys from the responses of their local PUF duplicate to these challenges.  In this sense, the protocol directly follows the traces of earlier PUF-based key exchange protocols \cite{TS,PhysicalTM,Brzuska} --- with the major advantage that no physical objects need to be exchanged between users for each protocol run, and that keys can be derived by an entire group of users, not just pairs of users.

Overall, the situation is illustrated in Fig. \ref{fig:Network}.  Interestingly, our approach does not require the group members to permanently store digital secrets \cite{Ruh2022} such as, for example, secret cryptographic keys or secret CRP-lists of a PUF as in traditional approaches. Conceptually, such vulnerable permanent storage of a digital key is replaced by the physical possession of a PUF duplicate \cite{Ruh2022}).  
This has various security upsides:  As the (necessarily physical) theft of a PUF cannot be accomplished merely via a digital communication channel or malware access, and will not go unnoticed.  Furthermore, also physical attacks on the PUF duplicates appear far more difficult than simple physical extraction of digital keys.  This uplifts security against invasive and malware attacks to a qualitatively new level \cite{Ruh2022}.  

Another promising application of duplicate PUFs is as PUF-based ciphers. 
It is a known idea to use PUFs as the secret round function in a Feistel-structure block cipher \cite{AMSST2010}.
An interesting aspect of this construction is that the input-output behaviour of the PUF does not need to be inverted for decryption.
The PUF is challenged in exactly the same way during both encryption and decryption,
but the helper data is handled differently:
during encryption helper data is generated from the PUF output and stored as part of the cipher text;
during decryption the helper data is used to error-correct the PUF output.
Due to the unclonability of standard PUFs such a cipher can be used only on a single device,
e.g. for memory encryption.
The introduction of PUF duplicates now allows us to extend this concept to multiple devices with identical encryption/decryption functionality.

As final example, duplicate PUFs also allow the forgery-proof {\it ``labeling''} or {\it ``tagging''} of classes of valuable objects, and can thus act as effective anti-counterfeiting mechanism.  In this approach, all items or objects belonging to a given class (e.g., all banknotes with a certain value, all products from a certain category or brand, etc.) are labeled consistently with identical duplicates;  other classes have their own duplicates, with a different inner construction plan.  It is obviously of particular relevance here that the duplicates are mass-producible, while still remaining unclonable for external adversaries.  Interestingly, if standard PUFs were used for the same application, then the individual properties of each single PUF would need to be measured at production time, stored in the form of PUF CRP lists on a central server, and a connection to this central server is required in each verification of a given label.  PUF duplicates could  overcome these practical issues.   
Please note that the  international trade in counterfeited products is estimated around 1 trillion Dollars per year, with numbers still growing \cite{trillion}.  This creates a pressing need for simple, yet practical item tagging methods,  such as our unclonable duplicates.

\section{Summary and Outlook}

Physical Unclonable Functions (PUFs) are commonly assumed to be unclonable by friends and foes, by their original manufacturer and external adversaries alike.  This paper showed for the first time that this key assertion is not necessarily met.  We demonstrated that PUF duplication by original manufacturers is practically achievable via latest fabrication techniques, for example by direct laser writing. 
Using such techniques, manufacturers may build many identical PUF duplicates from the same inner construction plan, while the fabricated PUFs can still remain unclonable for external adversaries.

The possibility of manufacturer-based PUF duplication has positive and negative implications at the same time.  On the one hand, it allows novel attacks:  This includes impersonation of unknowing users by malicious manufacturers in identification protocols \cite{Pappu2002}, or tacit cheating in more complex PUF schemes, for example key exchange or oblivious transfer \cite{Ruh2010,Ruh2013}. For most of these attacks, it is not even necessary that the fabricated duplicates remain unclonable for external adversaries;  the attacks even work without this special additional feature.  

On the other hand, PUF duplicates generated by physical one-way processes, in which the duplicates do remain unclonable for external adversaries, enable various new applications.  This includes key-free group identification and group message authentication over digital networks, or the key-free, forgery-proof labeling of valuable items.  High-quality duplicates also allow novel cryptographic encryption/decryption devices that operate without standard secret keys. In all cases, the key-free nature of the duplicates leads to novel and superior security features against malware and physical hardware attacks \cite{Ruh2022}.  

We believe that our work opens up manifold possibilities for follow-up activities.  This includes, first of all, implementing the abovementioned security applications in full detail in practice, together with identifying yet new ones.  Also the mathematical formalization of the two seminal concepts of physical one-way processes (POPs) and PUF duplicates appears very promising.  Finally, yet further structures, materials, and geometries for POPs and PUF duplication could and should be investigated.  Amongst others, the use of random and unclonable {\it ``stamps''} or {\it ``printing plates''}, similar as in existing money printing schemes, seems highly auspicious here.

\section{Methods}  \label{sec:methods}

We fabricate disordered scattering media as a collection of randomly-oriented polymer rods in a cubic volume. We employ an algorithm based on Jaynes' solution to Bertrand's paradox to ensure an on-average uniform filling of the volume and the random positioning of the rods in the cube~\cite{Marakis2019}.
The randomly intersecting polymer rods run from one facet of the cube to another, where we sorted the rod writing to start with those rods touching the glass substrate.
The samples used here are filled with 2000 rods in a cube of $15\,\mu$m lateral dimension, creating a dense volume packing, reducing the variations.

For the DLW we use the Nanoscribe professional GT tool \cite{Fischer2013}, where a scanning laser beam induces a physical change in a photoresist. The method allows the creation of almost arbitrary 3D nanostructures.
The samples were fabricated with a gel-like polymer photoresist ``IP-G'' from Nanoscribe, with a refractive index of 1.51 \cite{Gissibl2017}).
The high viscosity of the resist prevents drifting of the features in the samples during the writing processes, thereby minimizing deformation of the structures. The structures were fabricated by moving the piezo stage instead of using the faster galvo mirror mode in order to reduce  aberrations and further fabrication artifacts.
The illumination dose was set to be 
higher than the polymerization threshold with a laser beam power of 16.5\,mW and a piezo-motor scan speed of 200\,$\mu$m/s. 
This dose results in rods of an average thickness of 500\,nm, more than the 100\,nm that is achievable with a smaller illumination dose \cite{Guney2016,Seniutinas2018}. 
The slightly higher dose has two advantages: 1) thicker rods provide more robust mechanical support of the structure without any additional walls, thereby enabling self-supported structures unlike earlier demonstrations \cite{Deubel2004}.
2) Reduction of fabrication disorder through the complete cross-linking of polymer chains across the entire volume 
\cite{Lee2008}. Additionally, in the case of structures with a high filling fraction the higher energy dose reduces the pyramidal distortions that affect DLW structures, especially acrylic based photo-resists as IP-G.

\begin{figure}[hbt]
    \centering
    \includegraphics[width=0.5\textwidth]{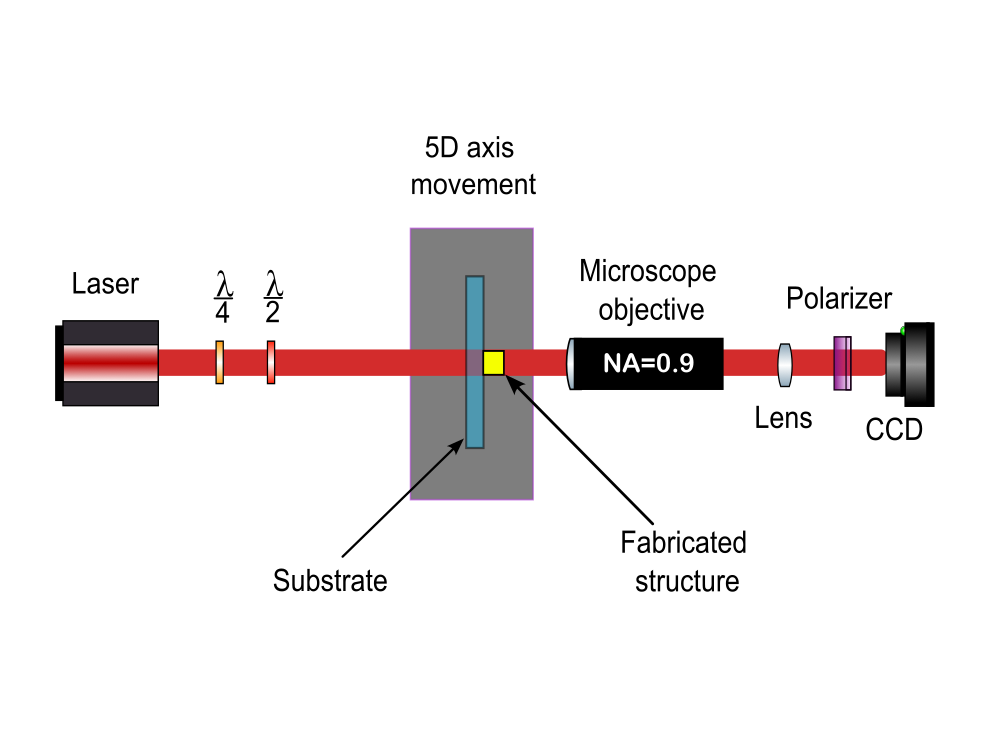}
    \caption{ Optical characterization set-up: A HeNe laser (left) is polarisation adapted by a quarter and a half-wave plate to illuminate the sample on a substrate  which is imaged onto a ccd camera via a microscope objective and a linear polariser.}
    \label{fig:setup}
\end{figure}

The speckle patterns in Fig. \ref{fig:bargraph}a) were taken by imaging a facet of the DLW sample while the opposite side was illuminated with polarized laser light. A microscopic set-up was assembled as shown in Fig. \ref{fig:setup}, where the sample was mounted on a sample holder with 2 axes angle positioning, while the holder is siting on a three-axes piezo translation stage. The beam was aligned to pass through the imaging objective with ${\rm NA}=0.8$ and imaged onto the imaging system. The laser beam has a Gaussian profile with diameter of $1\,$mm and could be considered a plane wave for the purposes of our experiment. Each structure was moved successively to the same location in front of the imaging objective.

\textbf{Acknowledgments}
E.M., R.U.\ and P.W.H.P.\ acknowledge financial support by the 
Nederlandse Wetenschaps Organisatie (NWO) via QuantERA QUOMPLEX (Grant No. 68091037), Vici (Grant No. 68047614) and NWA (Grant No. 40017607). U.R.\ and M.L.\ acknowledge support by the US Air Force Office of Scientific Research (AFOSR) via award FA9550-21-1-0039. 
B.\v{S}. acknowledges funding by Quantum Delta NL Groeifonds KAT-2. 
U.R.\ and M.L.\ acknowledge support by the US Air Force Office of Scientific Research (AFOSR) via award FA9550-21-1-0039. 
R.U.\ acknowledges support from the Office of Vice President for Research at the University of Iowa through the Jumpstarting Tomorrow award (003505). 

\medskip

\end{document}